\documentclass{article}
\usepackage[T1]{fontenc}
\usepackage[utf8]{inputenc}
\usepackage{ismir} % Remove the "submission" option for camera-ready version
\usepackage{amsmath,cite,url}
\usepackage{graphicx}
\usepackage{color}

\usepackage{tabularx}

\title{H2H Music Improv: A Communication Model and Audio-Visual Dataset for Music Improvisation}

\multauthor
  {Aleksandra Teng Ma$^{1,2,*}$ \hspace{0.5cm} Anthony Cammarota$^3$ \hspace{0.5cm} Jiayi Wang$^2$ \hspace{0.5cm} Alexandria Smith$^3$}
  {{\bf Cheng-Zhi Anna Huang$^1$ \hspace{0.5cm} Jeffrey Albert$^3$ \hspace{0.5cm} Alexander Lerch$^2$}\\
  $^1$ Human-AI Resonance Lab, Massachusetts Institute of Technology, USA\\
  $^2$ Music Informatics Group, Georgia Institute of Technology, USA\\
  $^3$ Creative Music Technology Lab, Georgia Institute of Technology, USA\\
  {\tt\small mtsandra@mit.edu}
  }

\def\authorname{A. T. Ma, A. Cammarota, J. Wang, A. Smith, C.-Z. A. Huang, J. Albert, and A. Lerch}

\usepackage[bookmarks=false,pdfauthor={\authorname},pdfsubject={\pdfsubject},hidelinks]{hyperref}
\usepackage{microtype}
\usepackage{paralist}
\usepackage{units}
\newcommand\blfootnote[1]{%
  \begingroup
  \renewcommand\thefootnote{}\footnote{#1}%
  \addtocounter{footnote}{-1}%
  \endgroup
}

\begin{document}

\maketitle
\blfootnote{$^*$ Work done while the author was at the Music Informatics Group, Georgia Institute of Technology.}
\begin{abstract}
Current real-time AI improvisation systems lack the communication awareness 
human musicians rely on: rather than treating communication as a foundational 
algorithm design concern, most systems layer interaction strategies post-hoc 
onto generative algorithms through explicit controls and predefined modes. 
This gap persists in part because no formalized, machine-readable communication model with musicians exists. To address this, we study how expert musicians communicate in free (non-idiomatic) 
improvisation, unconstrained by prior discussion or agreement. Through a collaborative co-design process with expert improvisers, we derive a communication model that (1) captures how free improvisers negotiate musical ideas and enter stable musical spaces, and (2) is formalized as a machine-readable annotation scheme. We further present the H2H (Human-to-Human) Music Improvisation dataset: six hours of audio-visual expert duo improvisations with clean per-player stems and per-player annotations of both their own intentions and their perception of their partner's intentions. To our knowledge, this is the first such dataset for free improvisation. Together, the communication model and the dataset offer a new lens and resource for studying musician communication and may in future inform the design of AI musical partners that communicate by design.

\end{abstract}
% \begin{abstract}
% In the machine learning community, there has been a lack of understanding and formalization of how musician communication unfolds during improvisation. Without such a communication model informing design, current real-time AI music improvisation agents naturally lack the communication awareness that human musicians rely on, even though they produce locally coherent output. To address this gap, we study how expert music improvisers communicate in free (non-idiomatic) improvisation, where no previous discussion constrains interaction. Through a co-design process in close collaboration with expert improvisers, we derive a communication model that (1) captures how free improvisers negotiate musical ideas and enter stable musical spaces, and (2) is formalized as a machine-understandable annotation scheme. We further present the H2H (Human-to-Human) Music Improv dataset: 6 hours of audio-visual expert duo improvisations with clean per-player stems and musician-provided annotations of both their own intentions and their perception of their partner's intentions. To our knowledge, this is the first such dataset for free improvisation. Together, these resources offer a new lens for studying musician communication and a foundation for building AI musical partners that listen and respond to humans, not merely generate.

% \end{abstract}

\section{Introduction}\label{sec:introduction}

Throughout history, improvisation has been a powerful force in the creation of new musical forms, as musicologist Ernst Ferand argued in his 1961 book, Improvisation in Nine Centuries of Western Music\cite{ferand_improvisation_1961}. Oftentimes, these new forms emerge from the interaction between self-organizing agents negotiating musical meaning in real time, as noted by Borgo\cite{borgo_sync_2022}. More recently, the advancement of artificial intelligence has opened new possibilities for innovation through musical improvisation. Yet most current machine learning systems train on existing music to generate "in the style of" human musicians, positioning AI as a surrogate rather than a source of innovation \cite{morreale_reductive_2025}. This is a design choice, not an inevitability. As Morreale et al. argue, machine learning abstracts musical structure in ways fundamentally different from human understanding, and this incommensurability holds the potential to unlock genuinely new creative territory \cite{morreale_reductive_2025}. Realizing this potential in improvisation, however, requires that humans and machine learning systems communicate meaningfully in real time.

% This is where current systems fall short. 
This human-machine communication incorporated at the algorithm design level is where current ML-based real-time improvisational systems fall short, even though they are capable of producing locally coherent and often surprising musical output. Most systems address it post-hoc through predefined interaction modes, such as trading or call-and-response\cite{thelle_spire_2021, blanchard_jam_bot_2025, brade_agents_2026, borg_somax_2022, borg_somax_2019}. These modes are categorical and static and do not reflect human musicians' natural communication, which unfolds as a temporal process.  
%This gap persists in part because there exists neither a formalized, machine-readable communication model, nor an annotated dataset that could be used for analysis or training. 

To address this, we examined musician communication in free, non-idiomatic improvisation as defined by Bailey \cite{bailey_improvisation_1993}, where music played by participating musicians is not tied to any pre-existing style and no prior discussion shapes the performance. While other forms of collaborative music making, such as jazz, also require participants to coordinate in real time\cite{keller_rhythm_2014}, free improvisation removes the scaffolds of agreed materials and requires that a significant part of the performance emerge from interaction and communication itself\cite{walton_improvisation_2015}.
% Other forms of collaborative music making also require musicians to coordinate and anticipate in real time~\cite{keller_rhythm_2014}. For example, a jazz combo or a chamber music ensemble does so over agreed material, form, and
% tempo, which scaffolds the coordination between players. However, free improvisation takes away the pre-existing materials, which means a significant part of the performance comes from interaction and communication itself \cite{walton_improvisation_2015}. 
Through a co-design process with expert improvisers,
we develop a communication model that formalizes how improvisers signal intent and communicate transition between musical states, expressed as a machine-readable annotation scheme. The co-design sessions also produced findings of their own.

We further present the H2H (Human-to-Human) Music Improvisation Dataset, containing six hours of audio-visual expert duo improvisations with clean per-player stems and per-player annotations, in which each performer annotates both their own musical intentions and their perception of their partner's. Beyond analysis of how improvised music evolves over time, the recordings support multi-modal work relating body and face gesture to music development, and tasks beyond improvisation such as multi-modal source separation. The dataset and an interactive preview are available at
\url{https://h2himprov.github.io}. This work serves as an initial empirical resource for studying how human musicians communicate and may inform the design of generative systems that incorporate communication at the algorithm level.

\section{Related Work}\label{sec:related-work}
\subsection{Modeling Communication in Co-Creative Systems}\label{subsec:h2m-systems}

Scholars have modeled communication with humans into their systems in different ways. Somax2 is an improvisation application developed at IRCAM that generates music through factor oracles trained on a specific corpus. Human and machine agents mutually influence each other via explicit parameter controls\cite{borg_somax_2019}. Spire Muse lets humans and the system negotiate through feedback buttons signaling approval or disapproval, triggering automatic switches in 
interaction strategies \cite{thelle_spire_2021}. jam\_bot fine-tunes transformer-based music language models on a specific performer's playing, offering interaction strategies that trigger generation at regular intervals, at every musical gesture, or upon explicit request \cite{blanchard_jam_bot_2025}. 
Across these systems, communication with the human musician is mediated through explicit controls and predefined interaction modes, which are mechanisms that are layered onto the generative algorithm rather than integrated into its design.

A second lineage treats the machine as an autonomous musical subject rather than a tool explicitly controllable. George Lewis's Voyager, first developed in the 1980s, holds its own musical values and communicates with human improvisers entirely through sound, in what Lewis calls a nonhierarchical, subject-subject model of discourse \cite{lewis_too_2000}. Heretic extends this philosophy by grounding its machine listening in a practitioner's framework: neural networks classify the human's playing into Anthony Braxton's Language Music categories, and Markov models select a response posture drawn from Joe Morris's postures of interaction \cite{brown_heretic_2019}. The Musebot ensemble was originally designed to be an autonomous ensemble where the human's role is not proactive\cite{eigenfeldt_collaborative_2015}, and later was adapted to incorporate interaction with humans\cite{brown_interacting_2018}. This lineage, however, remains largely rule-based; its communication-first philosophy has yet to be realized at the algorithm level in contemporary deep-learning systems.

\subsection{Human-to-Human Free Improvisation Behavior}\label{subsec:h2h-behavior}
% Prior work describes behaviors categorically, not temporally
% (Bailey, Borgo, Kuldkepp, Albert)
Free improvisation has been studied from ethnographic, psychological, and cognitive perspectives, each offering partial accounts of how musicians navigate collective performance. Bailey's foundational account\cite{bailey_improvisation_1993} establishes free improvisation as a practice defined by the absence of pre-existing style or form. Borgo situates this practice within complexity theory, arguing that musical form in free improvisation emerges from the interaction of self-organizing agents rather than from individual intention alone \cite{borgo_sync_2022}.

Empirical work has since sought to characterize this emergence more precisely. Canonne and Garnier asked expert free improvisers to perform in a collective free improvisation (CFI) setting and then qualitatively describe and segment their own playing, finding that musicians were defining sequences based on two orthogonal criteria: stability and desirability. They also identified four strategies used by free improv musicians: stabilization, "wait and see", "playing along", and densification\cite{canonne_cognition_2012}. In a follow-up study, they record improvisers using a MIDI pedal to mark significant changes in their own musical production in real time, and find that these individual decisions correlate with segmentation points identified by expert external listeners \cite{canonne_individual_2015}. Wilson and MacDonald, drawing on qualitative interviews with free improvisers, describe musical choices as an iterative cycle of maintaining or changing one's output, where changes take the form of either initiating new material or responding to a partner through adoption, augmentation, or contrast \cite{wilson_musical_2016}. Across these studies, accompanying recordings are limited or absent, and none model the communication exchange between performers as a structured, time-ordered process — a gap this work aims to address.
% \note{anything about datasets here?}\sandra{there's only one study where they released the dataset. i just added it. or were you thinking more like jazz improv dataset etc.?}\note{You have looked at related work from the co-creative perspective and from the model perspective. I was wondering what kind of related work might be out there from a dataset perspective.}

\section{Communication Model}\label{sec:model}
We formalize a communication model for free improvisation that is both grounded in the practice of expert musicians and operationalizable by computational systems. The communication model was developed through a co-design process by our interdisciplinary seven-member team with expertise in free improvisation, music information retrieval, machine learning, and human-computer interaction. %\note{if those are co-authors, can you call them collaborators?} \sandra{revised wording!}

\subsection{Co-Design Process}\label{subsec:codesign}
% Who, how, what emerged; thematic analyses of recordings
We adopted a co-design approach, where expert improvisers actively shape the design rather than only inform it~\cite{steen_co-design_2013, sanders_co-creation_2008}. Our co-design process iterated between playing sessions and interdisciplinary discussion, meeting bi-monthly over six months. The central question guiding the process was how musicians communicate musical ideas and traverse musical spaces over time during free improvisation. %To begin, we discussed\note{who is discussing what exactly?}\sandra{maybe we just take it out? i'll talk more about this in discussion} the scope of the study. 
Since our aim was to observe communication in its least constrained form, we imposed no pre-agreed material or musical concepts, including tempo, key, roles, and discussion of musical direction prior to performance. We also focused on duo configurations to isolate communication dynamics that become distributed and harder to trace in larger ensembles.

Having the scope laid out, we conducted pilot playing sessions with two expert improvisers, who also served as co-designers of the communication model. After each session, the improvisers reviewed the recording together and provided reflections on what had occurred: how ideas were introduced, how spaces developed or dissolved, and how transitions between sections came about. A recurring observation emerged from these reviews: the two musicians sometimes held different interpretations of  the same moment and sometimes there was a gap between produced intention and perceived intention. What one musician intended as a proposal of a new idea, the other sometimes perceived as natural continuation or even as a signal to move elsewhere. This asymmetry became a central design insight, and directly motivated our later decision to collect two sets of annotations in which each musician annotates their own intentions and their perception of their partner's intentions independently.

Through these review sessions, we also began to observe recurring temporal structures in how improvisations unfolded. There are periods of negotiation and searching, in which musicians put forward ideas without convergence. But there are also periods of relative stability, where the pair locked into a shared musical space, either building on the same idea or co-existing in parallel ideas. These observations were discussed among our experts from free improvisation, machine learning, and HCI, and then formalized into the communication model described in Sect.~\ref{subsec:formalization}. Then we recruited three additional expert improvisers to participate in recording sessions and annotate their own performances using the communication model. They were in agreement with the proposed model as it captures dynamics salient to free improvisation practice  broadly. We  discuss the limitations of this communication model during the annotation process in Sect.~\ref{subsec:limitations}.

All co-design sessions and interviews were recorded, transcribed, and subjected to 
thematic analysis\cite{braun_using_2006}; insights are discussed in Sect.~\ref{subsec:insights}.

\subsection{Formalization}\label{subsec:formalization}

\begin{figure}
    \centering
    \includegraphics[width=\columnwidth]{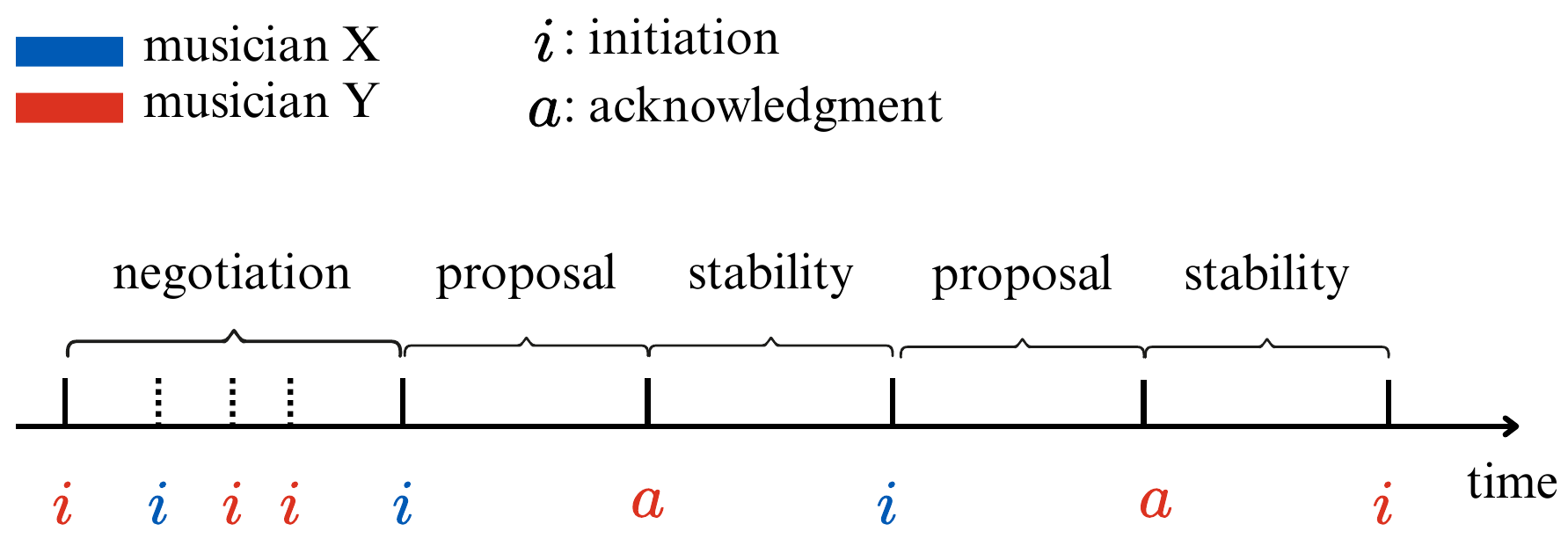}
    \caption{The communication model. Improvisers' interactions are modeled as a sequence of point-in-time actions: initiate (i) and acknowledge (a). Actions in time form 3 states: negotiation, proposal, and stability. Each improviser listens to the whole clip and provides timestamp markers annotating both their intended actions and their perception of their partner's actions.}
    \label{fig:comm_model}
\end{figure}

Our communication model represents the duo improvisation as a sequence of events on a one-dimensional timeline. The model is illustrated in Figure~\ref{fig:comm_model}. Conceptually, our communication model aligns with Canonne and Garnier's characterization of CFI as a coordination problem involving convergence on stable sequences and navigation of transitions between them \cite{canonne_cognition_2012}.

\subsubsection{Point-in-time Actions}
At any point%\note{this is a bit ambiguous -- does it mean at every time there is  an action or at any timr there could be an action; pretty sure it's the latter, but I misunderstood first}\sandra{revised wording!}
, a musician may perform one of two actions:

\begin{compactitem}
    \item \textbf{Initiate (i):} the introduction of a new musical idea or the signaling of a new sonic space. It may be the first sound of an improvisation, a deliberate shift in texture, or any gesture marking a departure from the current trajectory. Consecutive initiations may occur when one musician continues to put forward ideas not yet taken up by their partner.
    
    \item \textbf{Acknowledge (a):} a response from the partner that takes up a preceding initiation. By definition, an acknowledgment follows an initiation from the other musician; it does not initiate new material. Consecutive acknowledgments from the same musician are not defined, as acknowledgment is always directed at a specific preceding initiation.
\end{compactitem}
%\note{surely the musician can be also in different performance states than initiation or agreement? Also, is agreement an action? Maybe Acceptance? }

\subsubsection{States} 

Sequences of actions form three types of states, each characterizing a distinct mode of interaction:

\begin{compactitem}
    
    \item \textbf{Proposal:} the time between the start of an initiation and it receiving acknowledgment from the partner. We roughly define this as the proposal idea that is being built on. It usually follows negotiation or is the starting state, and is followed by stability from the initiating musician's perspective.
    
    \item \textbf{Stability:} the stretch of time following an acknowledgment during which both musicians develop the proposed idea, either by directly building on it together or by maintaining parallel streams that coexist within the same shared space. A stability state ends when a new initiation introduces a departure, returning to either negotiation or proposal. 

    \item \textbf{Negotiation:} a stretch of one or more initiations in search of alignment, none  met with acknowledgment. It usually follows stability or is the starting state.%\note{Would it flow better if you moved this to the third bullet instead of the first? I feel that this is a bit of a special state.}\sandra{yes}
\end{compactitem}

An improvisation is thus represented as an alternating sequence of states listed above. At annotation time, each musician provides timestamps marking actions and actors based on this model. 
%The formalization is kept coarse to represent navigation between musical spaces rather than the development of individual ideas, which keeps annotation tractable for musicians and yields a time-ordered event stream suitable for computational analysis. 

\begin{figure*}
    \centering
    \includegraphics[width=\textwidth]{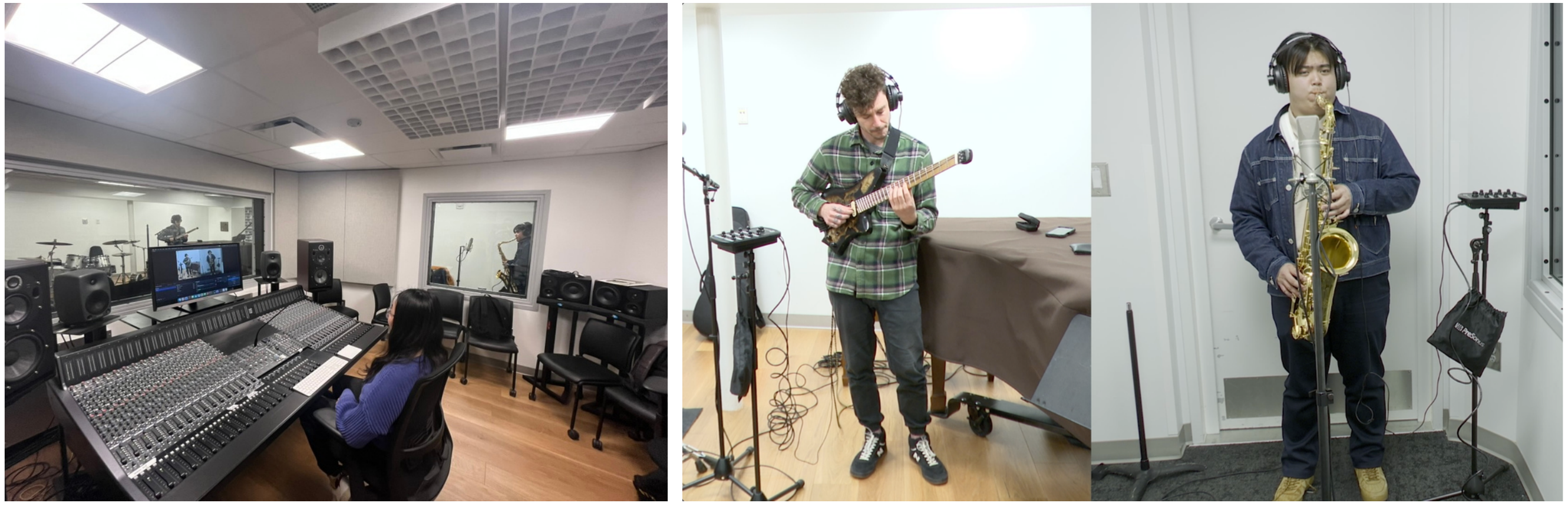}
    \caption{Left: studio setup with musicians in tracking and isolation rooms, connected via hearback system and inter-room window (not pictured). Right: frontal video recording of the improvisation.}%\sandra{delete \_fun in the file name for professional}
    \label{fig:studio_setup}
\end{figure*}
% Initiation/agreement timestamps -> negotiation/proposal/stability

\section{H2H Music Improvisation Dataset}\label{sec:dataset}

% To ground the communication model in empirical practice and facilitate related research, we collected the H2H (Human-to-Human) Music Improvisation Dataset: an audio-visual corpus of expert duo improvisations with clean per-player stems and per-player intention annotations. To our knowledge, this is the first free improvisation dataset with clean stems, synchronized video, and expert communication annotations, which is a unique contribution to research on computational analysis and modeling of improvisational interaction.

\subsection{Dataset Design}\label{subsec:dataset-design}

Three design priorities shaped the data acquisition process:
\begin{inparaenum}[(i)] 
\item capturing per-player audio without cross-bleed to support source-level analysis and generation,
\item preserving the naturalness of player interaction despite physical separation, and 
\item collecting intention annotations that reflect what individual musicians produce and perceive without interfering with the playing process.
\end{inparaenum}

\subsubsection{Recording Setup}

To maximize versatility for future research, we record per-player stems without bleed, which requires the two musicians to be in acoustically isolated rooms. To preserve natural interaction, we positioned them in the tracking room and isolation booth such that they maintain direct visual contact through a window and hear each other through a low-latency personal monitoring %\note{Is that a company name? I would just use generic terminology}
system (Figure~\ref{fig:studio_setup}). All 5 recruited musicians had extensive recording session experience and reported no impediment to interaction. Audio is recorded mono at \unit[48]{kHz}/\unit[16]{Bit} word length via Pro Tools and routed to OBS for synchronization with video. Video is captured by two synchronized SONY HXR-NX3 professional camcorders, each paired with a Blackmagic Design UltraStudio Mini Recorder capture device.

\subsubsection{Session Protocol}
Each session consisted of a pre-play interview, improvisation, and a post-play interview and annotation session. Musicians performed at intervals of their own choosing, accumulating about one hour of recorded play per session, with no pre-agreed material, tempo, key, or role discussed prior to playing.

\subsubsection{Annotation Protocol}
Following an improvisation session, each musician annotated
\begin{inparaenum}[(i)]
\item their own intentions and
\item their perception of their partner's intentions 
\end{inparaenum} by placing timemarks in Reaper.
This scheme is motivated by the occasional gap between produced and perceived intention observed during co-design. Annotation is performed post-hoc, as any in-session annotations, even automated ones, would interfere with the music itself. As our expert improviser co-designer put it, ``everything that you ask the player to do that's not simply playing the music is going to change the outcome of the music.''

Musicians were instructed to annotate only the initiations that signal the beginning of new musical spaces, not development within an established space. This granularity reflects our focus on how musicians navigate between ideas, as opposed to how they develop a single idea; it also keeps annotation tractable for musicians.
\begin{table}
\centering
\begin{tabularx}{\columnwidth}{llX}
\hline
\textbf{Musician} & \textbf{Time} & \textbf{Music Background} \\
\hline
guitar1    & 4h 3m   & 20+ years professional; active in jazz, free, commercial, and electro-acoustic improv\\
trombone2  & 3h 12m & 35+ years professional; active in jazz, free, and classical improv; credits on Grammy winning album\\
saxophone3 & 1h 57m  & 8 years pre-professional; active in jazz improv\\
trombone4  & 1h 1m  &  20+ years professional; member of internationally touring funk fusion ensemble\\
trumpet5   & 2h 4m  & 20+ years professional; active in jazz, free, and experimental improv; internationally performing artist\\
\hline
\end{tabularx}
\caption{Musician background and contribution time.}
\label{tab:musician_details}
\end{table}

\subsection{Dataset Statistics}

% The H2H Music Improvisation dataset comprises 37 clips of duo free improvisation, lasting about 6 hours and 8 minutes. It is recorded across 6 musician pairs, drawn from a pool of 5 expert improvisers. Each pair recorded approximately one hour of playing. In total, the dataset contains 1368 annotated actions across all clips and annotators. The 5 musicians' improvisational experience span diverse settings including jazz, classical, and commercial genres (funk, pop, rock, etc.), with 3 actively performing in free improvisation scenes. 4 out of 5 are professional performers with substantial recording and live performance experience, including credits on a Grammy-winning album and performance at major venues and festivals. Table~\ref{tab:musician_details} details each musician's background. The dataset features four instrument types: guitar ($\sim$4h), trombone ($\sim$4h across two players), saxophone, and trumpet (both $\sim$2h). Several musicians appear in multiple pairs, producing instrumental and interpersonal variety across the 6 configurations.
% Pair combination reflects a range of prior playing relationships: 2 pairs had performed together previously, while 4 pairs were first-time collaborators. Table~\ref{tab:musician_pair_stats} lists each pair's annotation count, duration, and prior relationship. This variability supports future analysis of how pre-existing musical familiarity shapes improvisational communication dynamics.

The H2H Music Improvisation Dataset comprises 37 clips of duo free improvisation totaling 6 hours and 8 minutes, recorded across 6 musician pairs drawn from a pool of 5 expert music improvisers (4 man-identifying and 1 woman-identifying, spanning from early-career to senior practitioners, all based in Atlanta). The musicians' experience spans diverse settings including jazz, classical, and commercial genres (funk, pop, rock, etc.), with 3 active in free improvisation. Four out of five are professional performers with substantial recording and live performance experience, including credits on a Grammy-winning album and performances at major venues and festivals. Table~\ref{tab:musician_details} details each musician's background. The dataset features four instrument types: guitar ($\sim$4h), trombone ($\sim$4h across two players), saxophone, and trumpet (both $\sim$2h).

Each pair recorded approximately one hour of playing. Pair combinations reflect a range of prior relationships: 2 pairs had performed together previously, while 4 pairs were first-time collaborators, allowing for future analysis of how musical familiarity shapes improvisational communication dynamics. Table~\ref{tab:musician_pair_stats} lists each pair's annotation count, duration, and prior relationship. In total, the dataset contains 1,368 annotated actions across all clips and annotators.

\begin{table}
\centering
% \footnotesize
\begin{tabularx}{\columnwidth}{Xccc}
\hline
\textbf{Pairing} & \textbf{Annotations} & \textbf{Duration} & \textbf{Played Together} \\
\hline
gtr1+trmb2    & 193 & 1h 6m & N \\
gtr1+sax3   & 200 & 54m   & N \\
trmb2+sax3 & 314 & 1h 3m & N \\
gtr1+trmb4    & 236 & 1h 1m & Y \\
trmb2+tpt5   & 246 & 1h 2m & Y \\
gtr1+tpt5     & 179 & 1h 2m & N \\
\hline
\textbf{Total}       & \textbf{1,368} & \textbf{6h 8m} & \\
\hline
\end{tabularx}
\caption{Dataset overview by musician pair.}
\label{tab:musician_pair_stats}
\end{table}

\subsection{Intention-Perception Alignment Analysis}\label{subsec:annotation}
\begin{figure*}
    \centering
    \includegraphics[width=\textwidth]{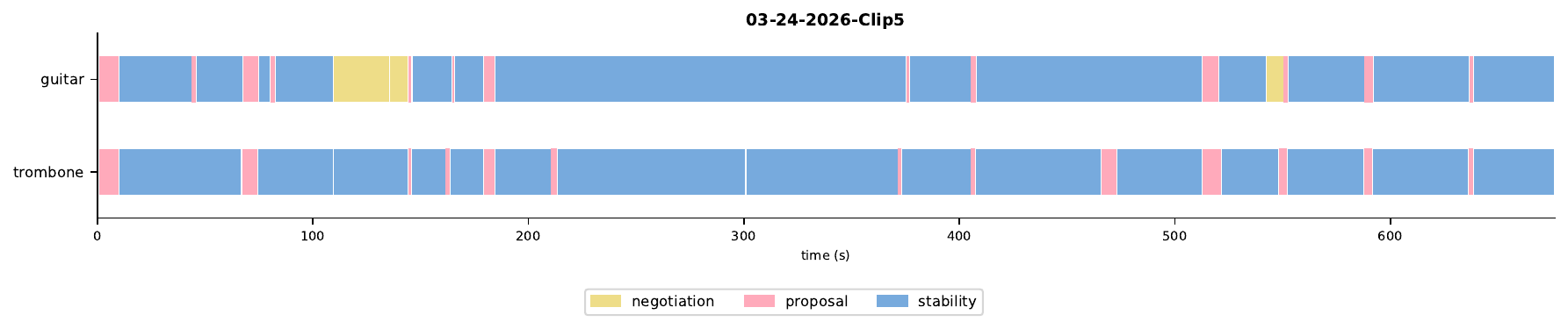}\vspace{-3mm}
    \caption{Example of paired annotations from both musicians on the same clip. 
Broad segmentation aligns, but musicians diverge on granularity and 
state assignment, illustrating the gap between produced and perceived intention.}
    \label{fig:align_example}
\end{figure*}

Our two-sided annotation scheme enables direct analysis of the gap between the musicians' intention and their perception. We examine this at both the clip and dataset level. 

\subsubsection{Individual Clip}

On a clip level, our musicians' annotation behavior mirrors a known challenge in composed music structure annotation: annotators may agree on the presence of structure while disagreeing on its hierarchical level~\cite{lerch_introduction_2022}. As illustrated in Figure~\ref{fig:align_example}, both musicians segment the improvisation into broadly similar regions, yet their annotations diverge in granularity. What one musician labels as a single stable state, the other may subdivide into finer states. Beyond hierarchy, we also observe cases where one musician perceived a space as negotiation while the other considered the proposal already accepted. These discrepancies reflect the inherent subjectivity of real-time musical communication, where intention and perception do not always align. 
%We encourage readers to explore an interactive preview of clips at \url{https://h2himprov.github.io} to develop their own intuitions about these patterns.

\subsubsection{Dataset Analysis}
To quantify the gap between intention and perception across the dataset, we compute two alignment metrics. Overall alignment is the fraction of the clip in which both musicians' annotations indicate the same state:
\begin{equation}
A_{\text{overall}} = \frac{\text{sec. both in same state}}{\text{sec. in clip}}
\end{equation}
Per-state alignment is the Intersection-over-Union (IoU) between the two annotators' markings for a given state:
\begin{equation}
A_{p} = \frac{\text{sec. both in state } p}{\text{sec. either in state } p}
\end{equation}
This per-state metric is robust to state-duration imbalance, which is relevant since stability states typically last substantially longer than negotiation or proposal states.

\begin{figure}
    \centering
    \includegraphics[width=\columnwidth]{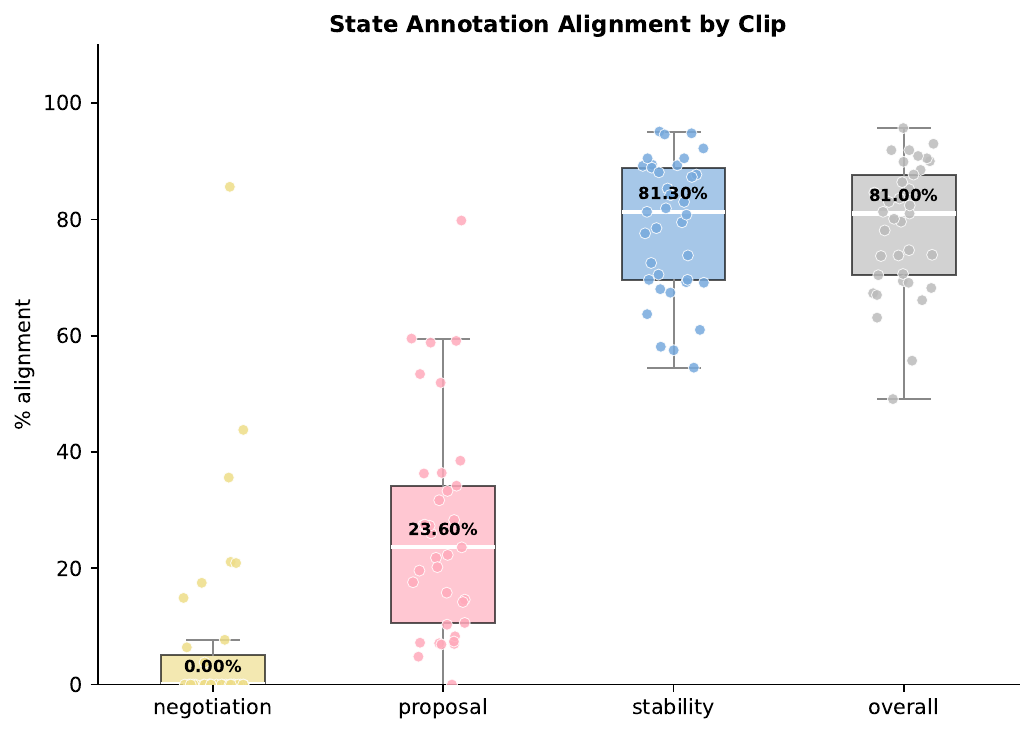}
    \caption{Overall and per-state alignment between paired annotations across 
    all clips (medians shown). Stability shows the highest agreement, while 
    proposal and negotiation reveal substantial divergence, not as annotation 
    error, but as evidence of genuine asymmetry between produced and perceived 
    intention.}
    \label{fig:state_align}
\end{figure}

Figure~\ref{fig:state_align} shows how much paired musicians agree on what state they are in. Stability shows high agreement (median 81.3\%), indicating that once musicians have settled into a shared space, both reliably perceive it as such. Proposal shows substantially lower agreement (median 23.6\%), reflecting how musicians frequently disagree about when or which initiation was actually taken up. Negotiation shows the most extreme disagreement with a median of 0\% (27 out of 37 clips have no negotiation overlap). This divergence is not an annotation error but evidence for the gap between perception and intention: one musician may experience themselves as putting forward unaccepted initiations, while the other does not register the interaction as negotiation. These findings empirically validate the co-design decision to collect two-sided annotations: intention and perception in improvisation are genuinely asymmetric, and a single annotator's perspective cannot capture this.

% For a music-in-music-out  co-creative machine partner, this means the system design needs to model bidirectional intent communication explicitly through feedback mechanisms, especially outside of stable states. %\note{I don't follow here. what do you mean with explicitly?}\sandra{like through feedback mechanisms instead of relying on humans to just pick up on the intentions}

% \subsection{Other Potential Use Cases}\label{subsec:usages}

% Beyond empirically grounding our communication model in practice, the H2H dataset supports several downstream research directions. Its multi-camera synchronized recordings with clean per-player stems extend existing source separation resources such as MUSIC\cite{zhao_sound_2018}, URMP\cite{li_creating_2019}, MUSDB\cite{rafii_musdb18_2017, rafii_musdb18-hq_2019}, and MoisesDB\cite{pereira_moisesdb_2023}. The videos enable analysis of the relationship between body and facial gestures and musical development. The dataset also supports more empirical investigation of music improvisation itself: how musical ideas are proposed and evolved within a stable space, and what gestures musicians use to signal transitions into a new one, as well as broader questions about interaction dynamics between improvisers.
% \note{Not sure about this subsection. I could see this both in the introduction and conclusion.}

\section{Discussion}\label{sec:discussion}

\subsection{Insights from the Co-Design Process}\label{subsec:insights}
We transcribed recordings from the co-design discussions, including pre- and post-play discussions, and conducted thematic analyses. Selected quotes below illustrate key insights that shaped the study.
\subsubsection{What Musicians Listen for}
A recurring theme is that pitch is not the primary dimension musicians attend to when improvising with a partner. \texttt{guitar1} stated, "The texture is that immutable parameter. The pitches are mutable. You can swap them out for whatever." When asked to define texture, \texttt{guitar1} and \texttt{trombone2} listed note length, onset density, and rhythmic and melodic shape. Other musicians have mentioned energy (\texttt{trumpet5}), mood, timbre, tonality or lack thereof, and groove (\texttt{trombone4}), musical gestures of the other person (\texttt{saxophone3}) as the main reference. 

\subsubsection{How Musicians Signal Intention}

Musicians described signaling through musical content, silence, and at section boundaries mostly, visual cues. To propose a 
change, \texttt{guitar1} identified three mechanisms: increased 
energy, repetition with variation (``it has to continuously occur to 
trigger awareness in the other player''), and silence. Conversely, 
switching to a repetitive accompanying pattern signals endorsement: 
``I like what you're doing, keep going'' (\texttt{trombone2}). 
Silence reliably predicts a forthcoming initiation, although an initiation does not require prior 
silence. Since free improvisation has no predetermined endpoint, endings require their own signals. \texttt{guitar1} identified three types: a decrescendo fade, a sudden stop, and a rarer mutually felt ending with no obvious acoustic cue. Visual contact appears specifically at these moments, \texttt{trombone4} said, "If there was a cadence in the music, a sort of closure, I would look up and look at him."

\subsubsection{Role Dynamics}

Another recurring theme across sessions is the role of leading and following. It appears to operate 
along multiple independent axes simultaneously, shifting over the course of an improvisation. \texttt{guitar1} observed that \texttt{trombone2} was "leading in every sense except rhythm; in the rhythmic sense you [\texttt{trombone2}] are following 
me." \texttt{trombone4} 
similarly noted that \texttt{guitar1} frequently established the rhythmic space. \texttt{trombone2} further noted that role cannot be inferred from the sonic output alone: "the same sonic result could be leading sometimes and following sometimes. It doesn't define itself."

Social dynamics also shape role behavior. \texttt{trombone2} and 
\texttt{trumpet5} both described a tendency to defer to more experienced musicians in improvisational settings, suggesting that 
hierarchy in musical experience and personality can influence who initiates and who follows. This is partially reflected in our 
annotation data. Across all pairings, \texttt{guitar1} initiates consistently, averaging approximately 63\% of perceived initiations. \texttt{trombone2}, by contrast, shows more variable initiation behavior, and initiates most frequently when paired with \texttt{saxophone3}, who is a less experienced musician in the study.

\subsection{Limitations}\label{subsec:limitations}

First, the communication model is designed to capture interaction at the level of musical spaces and their transitions, and does not account for all interaction patterns observed in practice. Two edge cases arose during annotation sessions that the 
current model does not capture. One is rapid acknowledgment without uptake: a musician briefly responds to their partner's initiation but then returns to their own independent material (\texttt{trombone4}). The other is gradual convergence: rather than a discrete initiate or acknowledge event, two parallel streams slowly merge over time, "we each start to incorporate bits of what the other one was doing" (\texttt{trombone2}). Both patterns suggest that musical interaction is not always linear, and that the model's discrete event model may not fully capture the continuous, emergent nature of some improvisational exchanges.

Second, the last five recording sessions were each one hour long. Toward the end of these sessions, musicians reported a form of “creative fatigue” attributed in part to sustained exposure to the same musical environment (\texttt{trombone4}, \texttt{saxophone3}). Beyond session length, musicians also noted that their mental state on a given day independently affected their musical creativity and engagement (\texttt{trombone2}, \texttt{saxophone3}).

Finally, horn instruments in our dataset occasionally exhibit audio clipping due to their wide dynamic range that was hard to anticipate and adjust for in time. 

We also acknowledge that our pool of expert improvisers is small and not demographically representative: it skews male, is entirely formally trained, and all based in Atlanta, Georgia.

\section{Conclusion}\label{sec:conclusion}
% We presented a communication model for free, non-idiomatic duo improvisation, co-designed with expert improvisers and formalized as a machine-readable annotation scheme. We further introduced the H2H Music Improvisation Dataset, to our knowledge the first annotated audio-visual dataset of its kind with per-player stems.
We presented a communication model and the H2H Music Improvisation Dataset for free, non-idiomatic duo improvisation, developed through a co-design process with expert improvisers. The model formalizes improvisational communication and interaction as a sequence of initiation and acknowledgment actions composing into negotiation, proposal, and stability states, expressed as a machine-readable annotation scheme. The dataset is, to our knowledge, the first annotated audio-visual dataset for free improvisation; it provides six hours of recordings with clean per-player stems and per-player intention annotations.

Our analysis reveals that state alignment between paired musicians is high for stability but breaks down for proposal and negotiation, empirically validating the gap between perception and intention and the need for per-player annotation. Insights from the co-design process reveal that musician communication is richer and more socially situated than musical content alone can capture.

We hope these contributions serve as both an empirical resource and an invitation to rethink how communication can be incorporated at the design stage of an ML-based music generative algorithm, rather than layered on top, toward machines that truly communicate by design.

\section{Ethics Statement}
All musicians were compensated or credited for their participation. All participants provided informed consent and signed a public image release form. This study was approved by the Institutional Review Board (IRB).

\section{AI Usage Statement}
LLMs were used for the purpose of spellchecking and editing. An AI coding tool was used to build the preview and data website. 

% For BibTeX users:
\bibliography{references}

% For non BibTeX users:
%\begin{thebibliography}{citations}
% \bibitem{Author:17}
% E.~Author and B.~Authour, ``The title of the conference paper,'' in {\em Proc.
% of the Int. Society for Music Information Retrieval Conf.}, (Suzhou, China),
% pp.~111--117, 2017.
%
% \bibitem{Someone:10}
% A.~Someone, B.~Someone, and C.~Someone, ``The title of the journal paper,''
%  {\em Journal of New Music Research}, vol.~A, pp.~111--222, September 2010.
%
% \bibitem{Person:20}
% O.~Person, {\em Title of the Book}.
% \newblock Montr\'{e}al, Canada: McGill-Queen's University Press, 2021.
%
% \bibitem{Person:09}
% F.~Person and S.~Person, ``Title of a chapter this book,'' in {\em A Book
% Containing Delightful Chapters} (A.~G. Editor, ed.), pp.~58--102, Tokyo,
% Japan: The Publisher, 2009.
%
%\end{thebibliography}

\end{document}